\UseRawInputEncoding
\documentclass[%
reprint,
superscriptaddress, 
amsmath,
amssymb,
prb]{revtex4-2}
\usepackage[utf8]{inputenc}
\usepackage[T1]{fontenc}
\usepackage{graphicx}
\usepackage{orcidlink}
\usepackage{physics}
\usepackage{soul}
\usepackage{xcolor}
\usepackage{float}
\usepackage{chngcntr}
\begin{document}

\title{Coupling 4H-Silicon Carbide spins to a microwave resonator at milli-Kelvin temperature}

\author{\orcidlink{0009-0007-7498-0741}Ali~Fawaz}
\email{ali.fawaz@mq.edu.au}
\affiliation{School of Mathematical and Physical Sciences, Macquarie University, North Ryde, 2109, NSW, Australia
}

\author{\orcidlink{0000-0002-5667-7745}Jeremy~Bourhill}
\affiliation{Quantum Technologies and Dark Matter Research Lab, Department of Physics, University of Western Australia, Crawley, WA 6009, Australia}

\author{\orcidlink{0000-0002-8675-2291}Stefania Castelletto}
\affiliation{School of Engineering, RMIT University, Melbourne, 3000, VIC, Australia}

\author{\orcidlink{https://orcid.org/0000-0001-9659-8382}Hiroshi Abe}
\affiliation{National Institutes for Quantum Science and Technology (QST), 1233 Watanuki, Takasaki, Gunma 370-1292, Japan}

\author{\orcidlink{https://orcid.org/0000-0002-7850-3164}Takeshi Ohshima}
\affiliation{National Institutes for Quantum Science and Technology (QST), 1233 Watanuki, Takasaki, Gunma 370-1292, Japan}
\affiliation{Department of Materials Science, Tohoku University, Aoba, Sendai, Miyagi 980-8579, Japan}

\author{\orcidlink{0000-0002-3139-1994}Michael~Tobar}
\affiliation{Quantum Technologies and Dark Matter Research Lab, Department of Physics, University of Western Australia, Crawley, WA 6009, Australia}

\author{\orcidlink{0000-0001-9850-4992}Thomas~Volz}
\affiliation{School of Mathematical and Physical Sciences, Macquarie University, North Ryde, 2109, NSW, Australia
}

\author{\orcidlink{0000-0002-0257-4054}Maxim~Goryachev}
\email{maxim.goryachev@uwa.edu.au}
\affiliation{Quantum Technologies and Dark Matter Research Lab, Department of Physics, University of Western Australia, Crawley, WA 6009, Australia}

\author{\orcidlink{0000-0002-9081-5750}Sarath~Raman~Nair}
\email{sarath.raman-nair@mq.edu.au}
\affiliation{School of Mathematical and Physical Sciences, Macquarie University, North Ryde, 2109, NSW, Australia
}

\begin{abstract}
Coupling microwave cavity modes with spin qubit transitions is crucial for enabling efficient qubit readout and control, long-distance qubit coupling, quantum memory implementation, and entanglement generation.
We experimentally observe the coupling of different spin qubit transitions in Silicon Carbide (SiC) material to a 3D microwave (MW) resonator mode around 12.6~GHz at a temperature of 10~mK.
Tuning the spin resonances across the cavity resonance via magnetic-field sweeps, we perform microwave cavity transmission measurements.
We observe spin transitions of different spin defects that are detuned from each other by around 60-70~MHz.
By optically exciting the SiC sample placed in the MW cavity with an 810~nm laser, we observe the coupling of an additional spin resonance to the MW cavity, also detuned by around 60-70 MHz from the centre resonance.
We perform complementary confocal optical spectroscopy as a function of temperature from 4~K to 200~K, using a part of the same sample used for the cavity measurements.
Combining the confocal spectroscopy results and a detailed analysis of the MW-resonator-based experiments, we attribute the spin resonances to three different paramagnetic defects: positively-charged carbon antisite vacancy pair (CAV$^+$), and the negatively-charged silicon vacancy spins located at two different lattice sites, namely V$_1$ and V$_2$ spins.
The V$_1$ and V$_2$ lines in SiC are interesting qubit transitions since they are known to be robust to decoherence.  Additionally, the CAV$^+$-transition is known to be a bright single-photon source. Consequently,  the demonstration of the joint coupling of these spin qubits to a MW cavity mode could lead to interesting new modalities: The microwave cavity could act as an information bus and mediate long-range coupling between the spins, with potential applications in quantum computing and quantum communication, which is an especially attractive proposition in a CMOS-compatible material such as SiC.
\end{abstract}


\maketitle

\section{Introduction}
An ensemble of solid-state spin qubits with long coherence times, when coupled to a microwave resonator at milli-Kelvin (mK) temperatures, can serve as a quantum memory for superconducting (SC) qubits \cite{julsgaard2013quantum,PhysRevLett.105.140501,PhysRevX.4.021049,grezes2016towards}. Such hybrid architectures are interesting as they address one of the main bottlenecks in SC-based quantum computing: the relatively short coherence times of superconducting qubits, which limit large-scale information storage and transfer \cite{PhysRevLett.105.140503, PhysRevLett.107.220501,PhysRevLett.105.140502,PhysRevB.92.014421}. By providing a long-lived memory, hybrid systems involving spins could enable enhanced quantum error correction capability and quantum computing  \cite{PhysRevA.111.012621,PhysRevLett.103.070502}.

Experimental coupling to microwave resonators at mK temperatures with different spin qubit ensembles, such as nitrogen-vacancy spins in diamond (NV) \cite{amsuss2011cavity, angerer2016collective}, P1-spins in diamond \cite{creedon2015strong}, rare-earth ion spins \cite{bushev2011ultralow, probst2014three, PhysRevB.103.214305}, relevant for the application mentioned above, are reported in the literature. 
Spin quantum systems in SiC are emerging as strong contenders for quantum technology applications, such as quantum sensing and quantum networks, due to their robust spin coherence even at room temperature and the potential for integration into a CMOS-compatible platform \cite{castelletto_silicon_2020}. A key advantage of SiC over other spin host materials is its mature micro-fabrication ecosystem \cite{koehl_room_2011,zetterling_process_2002}, which positions SiC spin-based technology to accelerate the development of scalable, real-world quantum devices.

Silicon-vacancy (V$_2$) spins in SiC have been coupled to microwave resonators at temperatures ranging from ambient to 4 K \cite{gottscholl_superradiance_2022,gottscholl_room-temperature_2023,fischer_highly_2018}, enabling continuous-wave masers and microwave amplifiers. However, to the best of our knowledge, no studies have reported the coupling of any SiC spins to microwave resonators at millikelvin temperatures. Exploring this temperature regime is interesting because spin coherence times are generally enhanced at lower temperatures, enabling access to narrower spin linewidths and stronger coherent interactions \cite{brereton_spin_2020, PhysRevB.95.045206}. In addition, the mK regime is in the operating range of SC quantum circuits, including flux qubits and high-quality factor microwave resonators, making it possible to integrate SiC spin systems with SC quantum technologies for hybrid quantum devices \cite{PhysRevLett.105.140501, doi:10.1021/acsnano.0c03167}

Here we study the coupling of thermally intialized spins in 4H-polytype SiC to a microwave resonator mode around 12.6 GHz at mK temperatures.
We also probe the coupled spin cavity system by applying non-resonant optical excitation to the spins.

\section{Experimental platform}

\begin{figure}
\centering
\includegraphics[width=\linewidth]{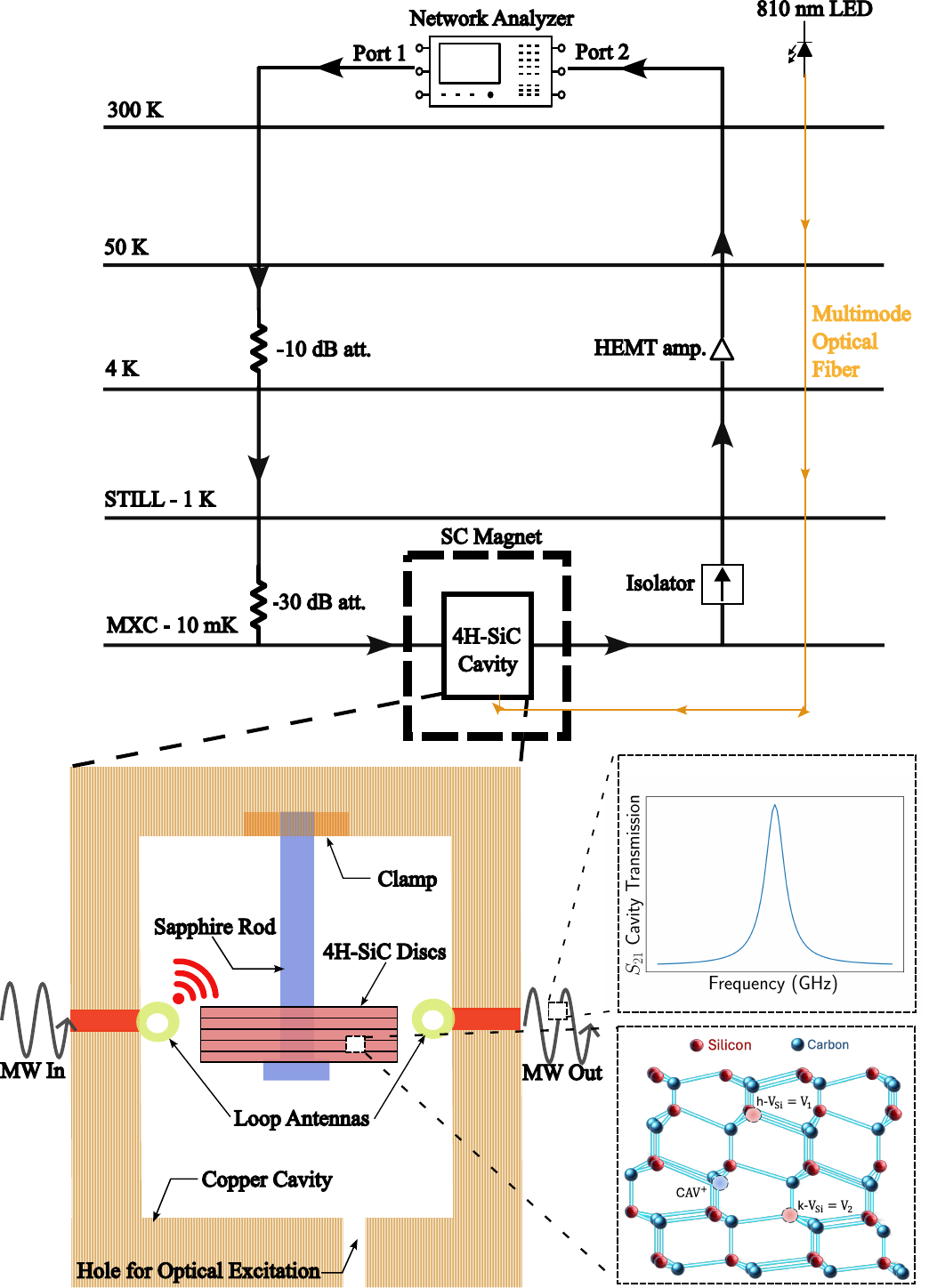}
\caption{Schematic diagram of microwave spectroscopy experimental setup in dilution refrigerator at 10~mK, showing 4H-SiC discs inside a copper cavity containing a sapphire rod acting as a secondary MW resonator. MXC refers to the $^3$He/$^4$He mixing chamber of the dilution refrigerator. The cavity is situated inside the bore of a SC magnet and cooled to 10~mK. Loop antennas couple microwaves inside and outside the cavity through which microwave S$_{21}$ parameter transmission measurements were performed. A hole at the base of the cavity is used to illuminate the SiC discs. The crystal lattice structure of 4H–SiC is also shown, showing CAV$^+$, V$_1$ (h site) and V$_2$ (k site) silicon vacancy defect sites.}
\label{fig:experimentalsetup}
\end{figure}

Our experimental platform is similar to the one described in reference \cite{hartnett_microwave_2011}, and is schematically depicted in Fig. \ref{fig:experimentalsetup}. 
The platform consists of a cylindrical MW cavity made from oxygen-free copper (height 24.13 mm, diameter 19.85 mm).
The oxygen-free copper enhances the cavity quality factor due to its high electrical conductivity and minimal concentration of magnetic impurities \cite{wang_operations_2022}.
The copper cavity has a top lid that can be fixed using appropriate screws and a 1-mm hole at the cavity base where an optical fiber cable can be attached for optical excitation.
A sapphire rod with a length of 20 mm and with two different cross-sectional radii (a long, thin rod with a thick disc at one end) was affixed to the cavity lid. 
The sapphire rod holds the SiC materials containing spins inside the cavity and also acts as a secondary dielectric resonator inside the copper cavity. 
Two loop antennas, oriented at 45 degrees on opposite cavity walls, couple microwave signals in and out of the cavity, enabling efficient excitation and detection of both transverse electric (TE) and transverse magnetic (TM) cavity field modes.
The entire platform is housed inside the bore of a 3-T SC magnet (American Magnetics Inc.), which itself sits inside a dilution refrigerator (Bluefors LD), enabling experiments at mK temperatures. 

\begin{figure*}
\centering
\includegraphics[width=\linewidth]{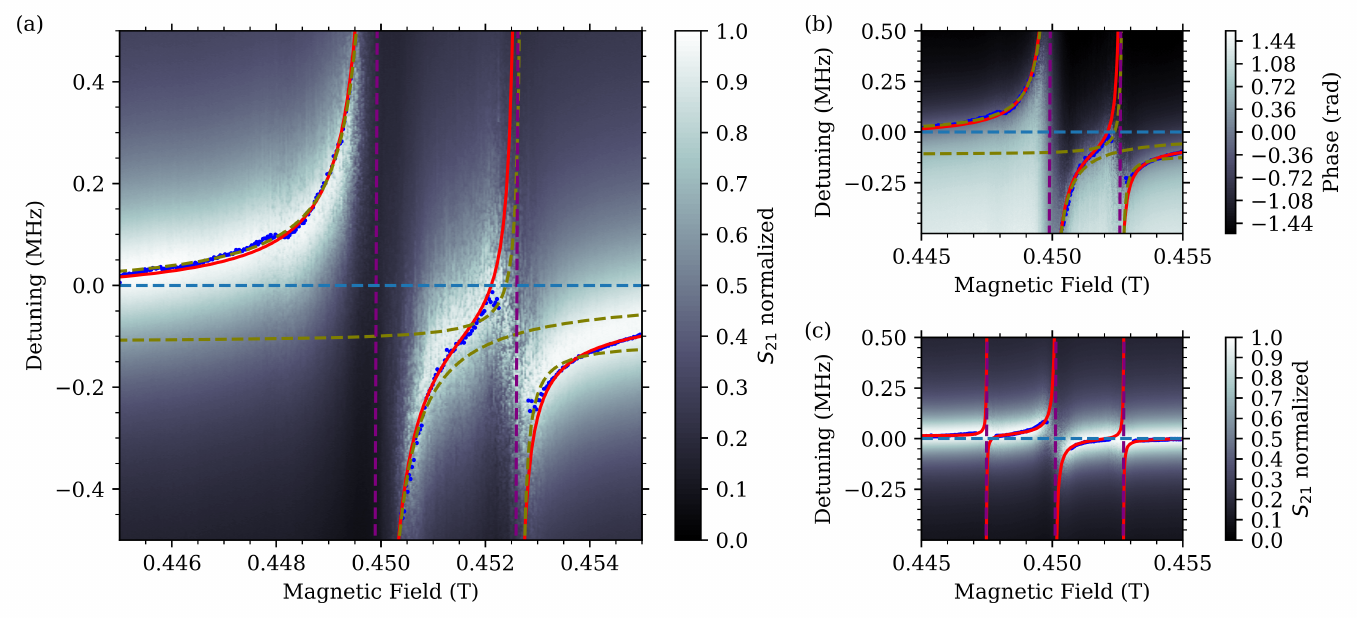}
\caption{Microwave S$_{21}$ transmission data obtained at a temperature of 10~mK: (a) Normalized S$_{21}$ as a function of detuning from the cavity resonance and axial magnetic field, showing two big avoided crossings. Blue dots represent frequency points where transmission is maximum based on S$_{21}$ fitting (see Appendix \ref{AppendixB}). The olive-dashed curve shows a fit with the uncoupled equation Eq. (\ref{eq:f1}), showing disagreement with experimental data. This is notable in the avoided crossing on the right side, where it is shifted from the fundamental cavity resonance frequency (blue dashed line) due to the interaction with the hybrid mode of the avoided crossing on the left. The red curves shown are fitted curves with Eq. (\ref{eq:f1}) and (\ref{eq:f2_c}). The horizontal dashed line shows the pure cavity resonance frequency at 12.59345~GHz. Purple dashed lines show fitted spin resonance lines $\omega_s=\gamma_eB-2D$ for each transition. The fitted values are ($g_1/2\pi=2.50\pm0.01$~MHz, $D_1/2\pi=2.01\pm0.10$~MHz) for the left avoided crossing and ($g_2/2\pi=1.34\pm0.02$~MHz, $D_2/2\pi=39.76\pm0.14$~MHz) for the right avoided crossing. (b) S$_{21}$ phase as a function of detuning from the cavity resonance and axial magnetic field, overlayed with fitted curves from (a). (c) S$_{21}$ measurement repeated with 810~nm optical excitation ($P \sim 20$~$\mathrm{\mu W}$). A new feature emerges near B=0.477~T, and the avoided crossings are reduced in size. The fitted values of the avoided crossings from left to right are ($g_3/2\pi=0.37\pm0.01$~MHz, $D_3/2\pi=-31.85\pm0.45$~MHz), ($g_1/2\pi=0.96\pm0.01$~MHz, $D_1/2\pi=5.08\pm0.07$~MHz) and ($g_2/2\pi=0.41\pm0.01$~MHz, $D_2/2\pi=41.55\pm0.47$~MHz).}
\label{fig:results}
\end{figure*}

The SiC consists of three complete discs ($11.85 \pm 0.1$~mm diameter, $375\pm10$~µm thickness) and two partial 4H-SiC discs (some parts broken off from two of the five full discs (approx. 7\% of total volume), and the broken pieces were not inside the microwave resonator).
Each disc has a central 2~mm diameter hole for secure mounting to the sapphire rod.
The SiC discs were electron irradiated with an energy of 2 MeV and fluences of 1 $\times 10^8$ cm$^{-2}$ to enhance the concentration of vacancies related paramagnetic defects, primarily the $V_{\rm{Si}}$.
The same SiC discs were used in reference \cite{hartnett_microwave_2011} before irradiating the sample to create spin defects, and all the discs were in complete form in reference \cite{hartnett_microwave_2011}. The material was originally sourced from the CREE corporation (CREE part number W4TRD0R-0D00).
Using finite element modelling in {\ it COMSOL Multiphysics}, we identify that the cavity can host resonances at different frequencies above 10 GHz; in particular, it can host whispering gallery modes (WGM) due to the SiC disc shape (See Appendix \ref{appendixA} for details).

The spin transitions in the SiC material can be brought to resonance with the resonances of the MW resonator by tuning the spin transitions via an external magnetic field parallel to the spin quantization axis that induces a Zeeman splitting.
We control the spin resonance using the magnetic field of a 3~T SC magnet.
The magnet is controlled by an AMI430 magnet controller, enabling precise magnetic-field tuning of the spin resonance across the cavity resonance.
In our system, the magnetic field is applied along the crystal's c-axis, which coincides with the quantization axis of silicon vacancy spins \cite{nagy_quantum_2018}. This field induces a Zeeman splitting of the spin states, separating the spin projections aligned with the field. As a result, only the spin components parallel to the c-axis (i.e., along the magnetic field direction) contribute to the observed coupling and dynamics.
We performed MW resonator transmission measurements ($S_{21}$) by sending a MW probe signal through the resonator.
For this, we used a Keysight N5234B PNA-L network analyser, operating from 10~MHz to 43.5~GHz and delivering -30 dBm to -20 dBm of output power, in combination with cryogenic attenuators (-40~dB total attenuation) for optimising input signal quality and minimising noise. 
An isolator on the output side of the cavity prevented reflections, and a 4K HEMT amplifier amplified the output signal for detection.
We first experimentally characterized five resonances of the filled resonator at 4 K (see Appendix \ref{AppendixB}), with the resonance at 12.6 GHz displaying the highest $Q$ factor and the clearest spin signatures; thus, we focused on this particular resonance for the rest of the experiments. 
At 10~mK, we performed S$_{21}$ transmission measurements on this mode to fit the cavity transmission at zero magnetic field and determine the cavity frequency $\omega_c/2\pi = 12.59345$ GHz, linewidth $\kappa/2\pi = 0.22$ MHz, and quality factor $Q = 57,232$. The Q-factor of this mode was found to differ at 10~mK and 4~K (Q~$\sim$~75000): The measurements at these two temperatures were performed on separate occasions, and the samples were repositioned between the measurements and could have therefore contributed to the change in Q-factor as well as the resonance frequency. 
Details on S$_{21}$ fitting methodology are provided in Appendix \ref{AppendixB}.
Furthermore, we carried out $S_{21}$ measurements by additionally applying a laser with a wavelength of around 810~nm (Thorlabs M810F2), delivered via an optical fiber to the SiC discs.

\section{Coupling SiC spins to the MW resonator at 10~mK}

We observe signatures of coherent coupling between the SiC spins and the microwave resonator mode near 12.6~GHz in S$_{21}$ transmission spectra as a function of magnetic field, shown in Fig.~\ref{fig:results}. In Fig.~\ref{fig:results}(a) and (b), the transmission amplitude and phase both display two prominent avoided crossings around B$_1 = 0.4500$~T and B$_2 = 0.4525$~T, indicating hybridization between the spin and cavity photon eigenstates. These polariton modes confirm strong coupling in the system. The phase response, Fig.~\ref{fig:results}(b), reveals near-$\pi$ phase shifts across resonances, highlighting the anti-crossings more clearly.
Figure~\ref{fig:results}(c) shows the S$_{21}$ response under 810~nm optical excitation at 20~$\mu$W, which introduces a new feature near B$_3 = 0.4477$~T and reduces the size of the previously observed avoided crossings.

To extract the spin parameters, we fit the observed avoided crossings using the standard coupled-mode expression~\cite{haroche_exploring_2013},

\begin{equation}
    \omega_{1, 2}^\pm = (\omega_c + \omega_s)/2 \pm \sqrt{g^2 + (\omega_c - \omega_s)^2/4}.
    \label{eq:f1}
\end{equation}

Here, $\omega_{1, 2}^\pm$ in Eq. (\ref{eq:f1}) gives the respective upper and lower branches of the avoided crossing, and the suffix 1, 2 refers to the first and second avoided crossing, respectively. $\omega_c$ is the bare cavity resonance frequency, and $\omega_s$ is the spin transition frequency tuned via the external magnetic field, following $\omega_s = \gamma_e B - 2D$ with $\gamma_e/2\pi = 28$~GHz/T~\cite{son_ligand_2019}. The parameters $g$, $\omega_c$, and $D$ are treated as fitting variables, where $g$ and $D$ denote the coupling strength and zero-field splitting, respectively.

When this model is applied to both avoided crossings simultaneously, we observe a deviation of the fit curve from the data due to the proximity of the two avoided crossings. The first hybrid mode influences the cavity response seen by the second spin and vice versa, similar to observations in other systems~\cite{ghirri_coherently_2016,zare_rameshti_indirect_2018,hyde_indirect_2016,zhang_nonreciprocal_2021}.
To account for this, we implement a modified fitting approach similar to reference \cite{zhang_nonreciprocal_2021} where $\omega_c$ is substituted with the hybrid mode frequencies and fitted iteratively. For the second avoided crossing, we substitute the expression for the lower polariton of the first mode using,
\begin{equation}
    \omega_2^{'\pm} = (\omega_1^-+\omega_s)/2\pm\sqrt{g^2+(\omega_1^--\omega_s)^2/4}.
    \label{eq:f2_c}
\end{equation}
Here $f^{'}_2$ is the frequency of the hybrid mode calculated using $\omega_1^-$ from Eq. (\ref{eq:f1}). Similarly, we fit the first crossing  $\omega_1^{'\pm}$ using the upper branch of the second using,
\begin{equation}
    \omega_1^{'\pm} = (\omega_2^++\omega_s)/2\pm\sqrt{g^2+(\omega_2^+-\omega_s)^2/4}.
    \label{eq:f1_c}
\end{equation}
This method provides improved agreement with experimental data. From the fits for the data presented in Figure~\ref{fig:results}(a), we obtain $g_1/2\pi=2.50\pm0.01$~MHz and $D_1/2\pi=2.01\pm0.10$~MHz for the first avoided crossing, and $g_2/2\pi=1.34\pm0.02$~MHz and $D_2/2\pi=39.76\pm0.14$~MHz for the second.
Fits to the data presented in Figure~\ref{fig:results}(c), using the same method above, give $g_3/2\pi=0.37\pm0.01$~MHz, $D_3/2\pi=-31.85\pm0.45$~MHz for the new feature, $g_1/2\pi=0.96\pm0.01$~MHz, $D_1/2\pi=5.08\pm0.07$~MHz for the central avoided crossing, and $g_2/2\pi=0.41\pm0.01$~MHz, $D_2/2\pi=41.55\pm0.47$~MHz for the rightmost avoided crossing.

\begin{figure*}
\centering
\includegraphics[width=\linewidth]{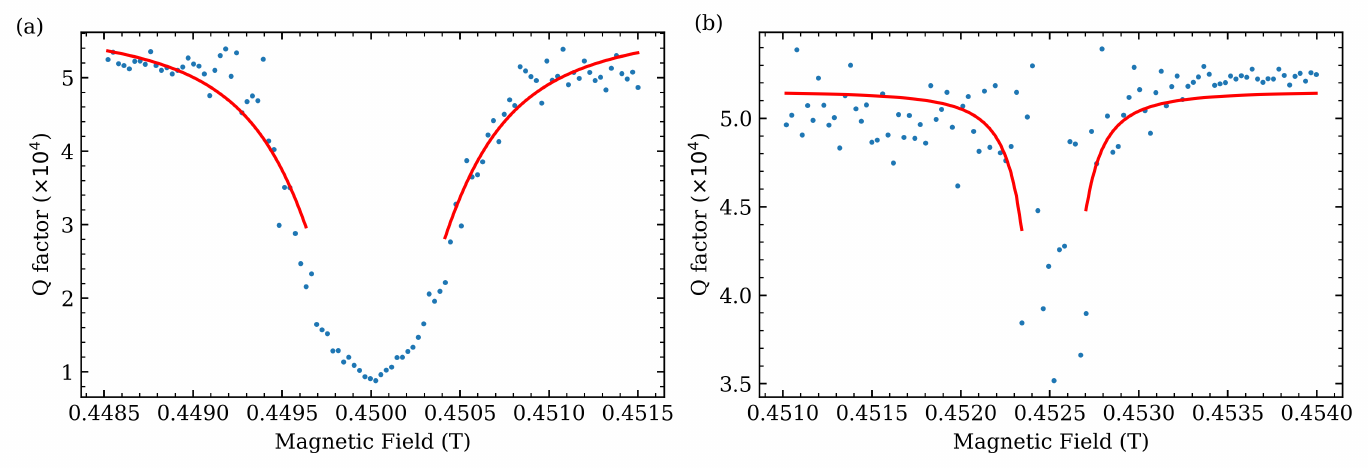}
\caption{(a) Q factor in the region of the first avoided crossing near B$_1$ fitted with Eq. (\ref{eq:Qfactor}) only in regions where $|\Delta/2\pi|\gtrsim10~\text{MHz}$. Blue points represent data points, and the red curve represents the fit. The spin linewidth extracted from the fit is $\Gamma_{d1}/2\pi = 4.27 \pm 0.19~\text{MHz}$. (b) Q factor in the region of the second avoided crossing near B$_2$ fitted with Eq. (\ref{eq:Qfactor}) only in regions where $|\Delta/2\pi|\gtrsim5~\text{MHz}$. The spin linewidth extracted from the fit is $\Gamma_{d2}/2\pi = 0.57 \pm 0.09~\text{MHz}$.}
\label{fig:Qfactorfitting}
\end{figure*}

In the vicinity of where the spin resonance approaches the cavity resonance, the Q factor of the resonator dips, as displayed in Fig. \ref{fig:Qfactorfitting}(a) and (b), for the avoided crossings near B$_1$ and B$_2$, respectively. The Q factor in these regions can be written in terms of the cavity resonance frequency $\omega_c$, spin linewidth $\Gamma_d$, cavity linewidth $\kappa$ and detuning $\Delta$, using a two-coupled harmonic oscillator model as \cite{PhysRevLett.105.140501,PhysRevA.102.013714}, 
\begin{equation}
    Q = \frac{\Delta^2+\left(\frac{\Gamma_d}{2}\right)^2}{g^2\Gamma_d + \kappa (\Delta^2 +\left(\frac{\Gamma_d}{2}\right)^2)}\omega_c.
    \label{eq:Qfactor}
\end{equation}
For the first avoided crossing, the fit in Fig. \ref{fig:Qfactorfitting}(a) is performed using known values of $g$, $\kappa$ and $\omega_c$, which are determined independently using the previously mentioned fitting of the avoided crossing data. 
The detuning is taken as $\Delta =\omega - \gamma_e B$. The spin linewidth $\Gamma_d$ and $\omega$ are left as free parameters. In addition, fitting is only performed in the region where the spin-cavity detuning $\Delta$ is much greater than the coupling $g$ in which Eq. (\ref{eq:Qfactor}) is valid. Only the regions beyond  $|\Delta/2\pi|\gtrsim10~\text{MHz}$ are included in the fitting for the first avoided crossing. For the second avoided crossing in Fig. \ref{fig:Qfactorfitting}(b), we fit only regions where $|\Delta/2\pi|\gtrsim5~\text{MHz}$ since the dip in the Q-factor occurs in a smaller magnetic field region, indicating that the spin linewidth is smaller in this region. We also note that the Q factor is modified due to the effects of the first avoided crossing. To fit, we include $\kappa$ as an additional free parameter for the second avoided crossing. From both fits, we extract the effective spin linewidths at each avoided crossing as $\Gamma_{d1}/2\pi=4.27\pm0.19~\text{MHz}$ and $\Gamma_{d2}/2\pi=0.57\pm0.09~\text{MHz}$, respectively. Since $g,\kappa, \Gamma_{d1},\Gamma_{d2}\textless \Delta$, fitting using Eq. (\ref{eq:Qfactor}) is valid.

\section{Discussion}
\begin{figure}
\centering
\includegraphics[width=\linewidth]{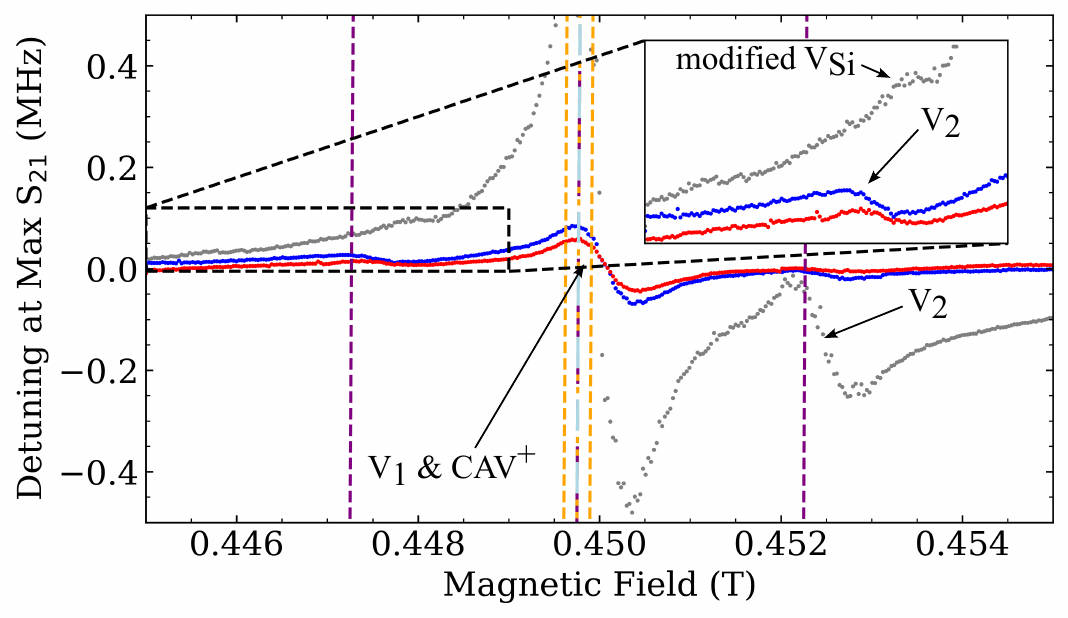}
\caption{Detuning from cavity resonance of maximum S$_{21}$ transmission as a function of magnetic field with optical pumping (blue represents 20~$\mu$W of optical power, red is 82~$\mu$W of optical power) and without optical pumping (grey). Points in the gap of the avoided crossings are also shown, but were not used to fit the data in Fig. \ref{fig:results}(a) and (c). The blue, orange and purple dashed lines represent theoretical spin transitions calculated using typical reported values of D=0 (CAV$^+$), D = 2~MHz (V$_1$) and D=35~MHz (V$_2$) and the spin-3/2 Hamiltonian $\hat{H}/\hbar=D S_z^{2}+\gamma_e \mathbf{B}\cdot\mathbf{S}$. Inset: close-up of the additional small avoided crossing arising from optical excitation. Also showing, modified V$_\text{Si}$ due to carbon antisites coupling to the cavity at 10~mK when no optical excitation is present (grey curve), more details are given in Appendix F.}
\label{fig:discussion}
\end{figure}
The detuning measured relative to the cavity resonance, where the S$_{21}$ transmission is maximum for cases with and without optical pumping (presented in Fig.~\ref{fig:results}), is compared in Fig.~\ref{fig:discussion}.
First, we attribute the resonance features to different spins in SiC, and then we will discuss the features in Fig.~\ref{fig:discussion}.

To get a better idea about the spin defects, we performed optical spectroscopy in a separate confocal microscopy setup from 200~K to 4~K on the complementary part of one of the incomplete SiC discs (a piece that was broken off from the sample used in the MW resonator measurement).
We observed zero-phonon lines (ZPLs) of the V$_1$ spins at 861~nm, V$_2$ spins at 917~nm, and positively-charged carbon antisite vacancy pair (CAV$^{+}$) spins with spectral emission ranging from 672~nm to 676~nm (see Appendix \ref{appendixC} for more details on the optical spectroscopy), which agree with previously reported ZPLs for these spins \cite{ivady_identification_2017,castelletto_silicon_2014}.
The complementary piece of SiC on which the optical spectroscopy was performed was part of the irradiated samples; therefore, we expect the SiC discs in the MW resonator to have identical spectral properties, i.e. to also contain these SiC spins.

Both V$_1$ and V$_2$ spins possess a spin-$3/2$ ground state, consisting of a total of four states with two degenerate ground states separated by $2D$, referred to as zero-field splitting (ZFS).
Typically reported ZFS values are $D/2\pi=2$ MHz for V$_1$ and $D/2\pi=35$ MHz for V$_2$ \cite{sorman_silicon_2000}. In contrast, CAV$^+$ spins are expected to have spin-$1/2$, which is associated with no zero-field splitting \cite{umeda_identification_2007}.
These theoretical $D$ values are close in value to those obtained from experimental data fits in the previous section.
The difference in the fit values and the theoretical values of $D$ may result from angular deviations of the c-axis to the magnetic field \cite{gottscholl_superradiance_2022}, local crystal‑strain fields  
that perturb the spin–spin Hamiltonian
\cite{breev_stress-controlled_2021,udvarhelyi_vibronic_2020}, small absolute‑field calibration errors in the
SC magnet
\cite{botsch_vectorial_2020}, or density‑dependent spin–spin correlations in
high‑concentration ensembles
\cite{brereton_spin_2020,lekavicius_orders_2022}.

We include the theoretically expected spin transition lines using the reported literature values of CAV$^+$, V$_1$ and V$_2$ for all possible dipole transitions in Fig.~\ref{fig:discussion}.
All these three features are within proximity to the spin transitions of CAV$^+$, V$_1$ and V$_2$.
The proximity of the CAV$^+$ and V$_1$ resonances at the cavity frequency implies that their features merge into a single avoided crossing, most likely due to ensemble-dephasing linewidths exceeding their frequency separation~\cite{kurucz_spectroscopic_2011}. 
In light of the above discussion, we attribute the resonance features observed in the microwave resonators to CAV$^+$ (at B$_1$), V$_1$ (at B$_1$), and V$_2$ (at B$_2$) spins.
Interestingly, the avoided crossing due to V$_2$ lies on the lower branch of the avoided crossing of V$_1$. This implies that in the region near the vicinity of B$_2$, both V$_1$ and V$_2$ are coupling to the cavity. Both spins share the same cavity mode, and therefore, one can expect indirect coupling between both spin species mediated by the cavity.

The additional resonance feature B$_3$ in the microwave cavity measurements with optical excitation present is consistent with an excited state transition of V$_2$ spins ($m_s=1/2\leftrightarrow3/2$), which under spin-optical polarization \cite{liu_silicon_2024} and/or (light-induced) thermalization is expected to be populated and therefore to enable coupling to the cavity mode. 
We also note that we observe smaller features corresponding to modified V$_\text{Si}$ coupling to the cavity \cite{davidsson_exhaustive_2022}. 
More of these modified V$_\text{Si}$ features, observed over a broader magnetic field range, are presented in Appendix F. Under optical excitation, however, these small features are no longer present.

Since the ensemble coupling follows $g = g_0 \sqrt{N}$, where $g_{0}$ and $N$ are the single spin coupling and number of spins, respectively, we can provide a lower bound to the number of spins for the maximum possible $g_{0}$ value involved in the coupling.
For this, we consider the coupling of SiC spins when there is no optical pumping present.
Assuming that the microwave magnetic field is linearly polarised along x-axis (where the spin quantisation axis is along z-axis) the quantized magnetic field can be written as $\mathbf{\hat{B}_x} = B_0(\mathbf{\hat{a}^\dagger}+\mathbf{\hat{a}})$, where $B_0 = \sqrt{\mu_0\hbar\omega_c/2V}$ \cite{Scully_Zubairy_1997}.
For the spin operator $\mathbf{\hat{S}}$ coupled to the microwave cavity, the interaction Hamiltonian between the spin and cavity can be expressed $\mathbf{\bar{H}}= g_e\mu_B \mathbf{\hat{B}_x}\cdot \mathbf{\hat{S}}/\hbar$. 
The single spin photon coupling can be expressed as $g_0 = g_e\mu_B B_{0} \bra{i}S_x\ket{f}/\hbar$, where the $i$ and $f$ denote the initial and final states of the spin transition. 
Since in our experiment, we have multi-spin species coupling and since we are estimating a lower bound to the number of spins involved in the coupling, we consider only the spin-3/2 system, as they are expected to have more single spin coupling strength.
The spin dipole transition element for spin-3/2 is $ \bra{-1/2}S_x\ket{-3/2} = \sqrt{3}/2$, given pure spin states and a linearly-polarized cavity field \cite{siegman1964microwave}. 
This leads to the single spin-photon coupling strength given as,
$g_0/2\pi = (g_e\mu_B/\hbar) \sqrt{(3\mu_0 \hbar \omega_c/8V)} \approx 0.12~\text{Hz}$.
Here we take the mode volume as the volume of the discs since we assume the field is primarily stored in the discs with the assumption that the measured 12.6~GHz mode corresponds to a whispering gallery mode, $V\approx 2.13\times 10^{-7} \text{m}^3$.
Hence, we estimate the minimum number of spins at the first avoided crossing $N_1=(g_1/g_0)^2\approx4.31\times10^{14}$ and the minimum number of spins at the second avoided crossing $N_2=(g_2/g_0)^2\approx 1.24\times10^{14}$.

Across all the observed resonance features, the cavity shows a consistent detuning trend, with positive detuning on the left and negative detuning on the right. This behavior indicates that the resonances arise from the net absorption of microwave photons by the spins, rather than any net gain supplied to the cavity. However, the strength of these features changes significantly under optical pumping. At an excitation power of 20~$\mu$W, the avoided crossings at B$_1$ and B$_2$ become weaker, while a new avoided crossing appears at B$_3$. Increasing the excitation power to 80~$\mu$W reduces all avoided crossings further compared to the 20~$\mu$W case. These observations show that the addition of the laser modifies the spin populations responsible for photon absorption at the different transitions.  Two possible causes can explain the changes: either thermal heating effects due to the laser or spin-selective optical pumping effects.

We first consider whether they can be explained by thermal effects alone. Assuming that the changes are governed purely by Boltzmann statistics, the reduction in coupling at B$_2$ under 20~$\mu$W optical excitation corresponds to an effective spin temperature of about 1.76~K. If heating were the only contributing effect, then the coupling at B$_3$ should be much weaker than at B$_2$, since the Boltzmann distribution predicts a smaller population difference for the $m_s = 1/2 \leftrightarrow 3/2$ transition (see Appendix E). However, our measurements show that the coupling strengths are similar (g$_2/2\pi = 0.41$~MHz, g$_3/2\pi = 0.37$~MHz). This discrepancy indicates that heating alone cannot account for the observed features.  

Spin-selective optical pumping provides a natural explanation for the appearance of the B$_3$ avoided crossing since silicon-vacancy spins in 4H-SiC are known to exhibit this property \cite{banks_resonant_2019,morioka_spin-optical_2022}. To evaluate its effectiveness, we estimate the optical pumping rate as $\Lambda_{exp} = \sigma_{abs} P \lambda/hcA$ where $c$ is the speed of light and $h$ is Planck’s constant. For the calculation, we use an absorption cross section of $\sigma_{abs} \approx 1 \times 10^{-17}$~cm$^2$, reported to be similar to NV centres in diamond \cite{hain_excitation_2014}, an optical power of $P = 20~\mu$W, a wavelength of $\lambda = 810$~nm, and an illuminated area of $A \approx 2 \times 10^{-5}$~m$^2$ (estimated from the divergence angle of light from the hole at the base of the cavity and the distance to the sample; the actual area may differ and we neglect any reflections that may occur inside the cavity). This gives $\Lambda_{exp} \approx 4 \times 10^{-3}$~Hz. For comparison, the longitudinal spin relaxation time of V$_2$ at 10~mK can be estimated to be about 9~s from reference \cite{simin_locking_2017}, corresponding to a decay rate of 0.11~Hz. Thus, the relaxation rate exceeds the optical excitation rate by a factor of about 27.5, implying that at any one time only a small fraction of spins is optically polarized. Therefore, while optical pumping may contribute to the observed changes and explains the appearance of B$_3$, its effect is expected to be weak on its own. 

The most consistent explanation is that the system exhibits a combined response from both thermal and optical effects. A portion of the spin ensemble is thermally redistributed according to the Boltzmann distribution, which reduces the coupling at B$_1$ and B$_2$, and is further influenced by the measured increase in the cavity Q factor under illumination (Appendix D), which reduces the effective spin population coupling to the cavity. At the same time, a small fraction of spins is optically pumped into non-thermal equilibrium populations, enabling the B$_3$ avoided crossing to appear with a strength comparable to that of B$_2$. Because the laser does not illuminate the sample uniformly, different spin sub-ensembles experience these effects to varying degrees, and the cavity transmission measurement reflects their averaged contribution.  

Additional support for this interpretation comes from measurements at 4~K, where thermal occupation is already broadened. At this temperature, even with much higher optical powers (on the order of mW), the avoided crossings change only slightly, consistent with the expectation that heating plays little role at elevated temperatures. Nevertheless, we still observe an enhancement of the avoided crossing feature that is equivalent to B$_3$ at mK temperatures with optical pumping. This behavior indicates that optical pumping modifies the spin populations, and at millikelvin temperatures, its effects compete with heating, which explains the reduction of the avoided crossing at B$_3$ with further increasing laser power.  

\section*{Conclusions}

In summary, we have demonstrated coupling of an ensemble of different spin species in 4H-SiC to a high-$Q$ 3D copper-sapphire microwave cavity at 10~mK. 
Using magnetic field-tuned transmission spectroscopy, we observe two well-resolved avoided crossings separated by 60-70~MHz, associated with ensembles of CAV$^+$, V$_1$ and V$_2$ spins. Our analysis, including fitting and optical spectroscopy, confirms the presence of these spin species, which are the only reported spin species in 4H-SiC to have ZFS that would allow them to be measured in the magnetic field region examined in this work. 
The observed coupling of V$_1$ and V$_2$ spins to the same cavity mode revealed cavity-mediated interaction effects through a shared hybrid spin-cavity mode. 
We also studied the effects of optical excitation with 810~nm light on these spin ensembles at mK temperature, and discussed the effects of heating and optical pumping on the spin-photon coupling.
Our results demonstrate the potential of SiC as a versatile host for hybrid quantum systems comprising multiple spin species, thereby paving the way for its application in scalable quantum technologies.

\begin{acknowledgments}
This research was supported by the Australian Research Council Centre of Excellence for Engineered Quantum Systems (EQUS CE170100009) and the ARC Centre of Excellence for Dark Matter Particle Physics (CE200100008). We acknowledge the EQUS Quantum Clock Flagship and a Centre Collaboration Award.
\end{acknowledgments}

\appendix

\renewcommand{\thefigure}{\Alph{section}\arabic{figure}}
\renewcommand{\theequation}{\Alph{section}\arabic{equation}}
\renewcommand{\thetable}{\Alph{section}\arabic{table}}
\counterwithin{figure}{section}
\counterwithin{equation}{section}
\counterwithin{table}{section}

\section{Finite-element simulations of cavity modes\label{appendixA}}
To identify and characterize resonant cavity modes, we performed finite-element simulations using COMSOL Multiphysics. 
These simulations were crucial for identifying particularly whispering-gallery modes (WGMs), enabled by the high permittivity ($\epsilon_{r}\approx9.3$ \cite{hartnett_microwave_2011}) of the 4H-SiC discs.
The cavity was modelled with perfect electric conductor boundaries, omitting minor structural features such as the optical hole and loop antennas. The sapphire rod and five complete 4H-SiC discs were considered in the simulation, simplifying the actual experimental configuration of two partial and three complete discs.

\begin{figure}[H]
\centering
\includegraphics[width=\linewidth]{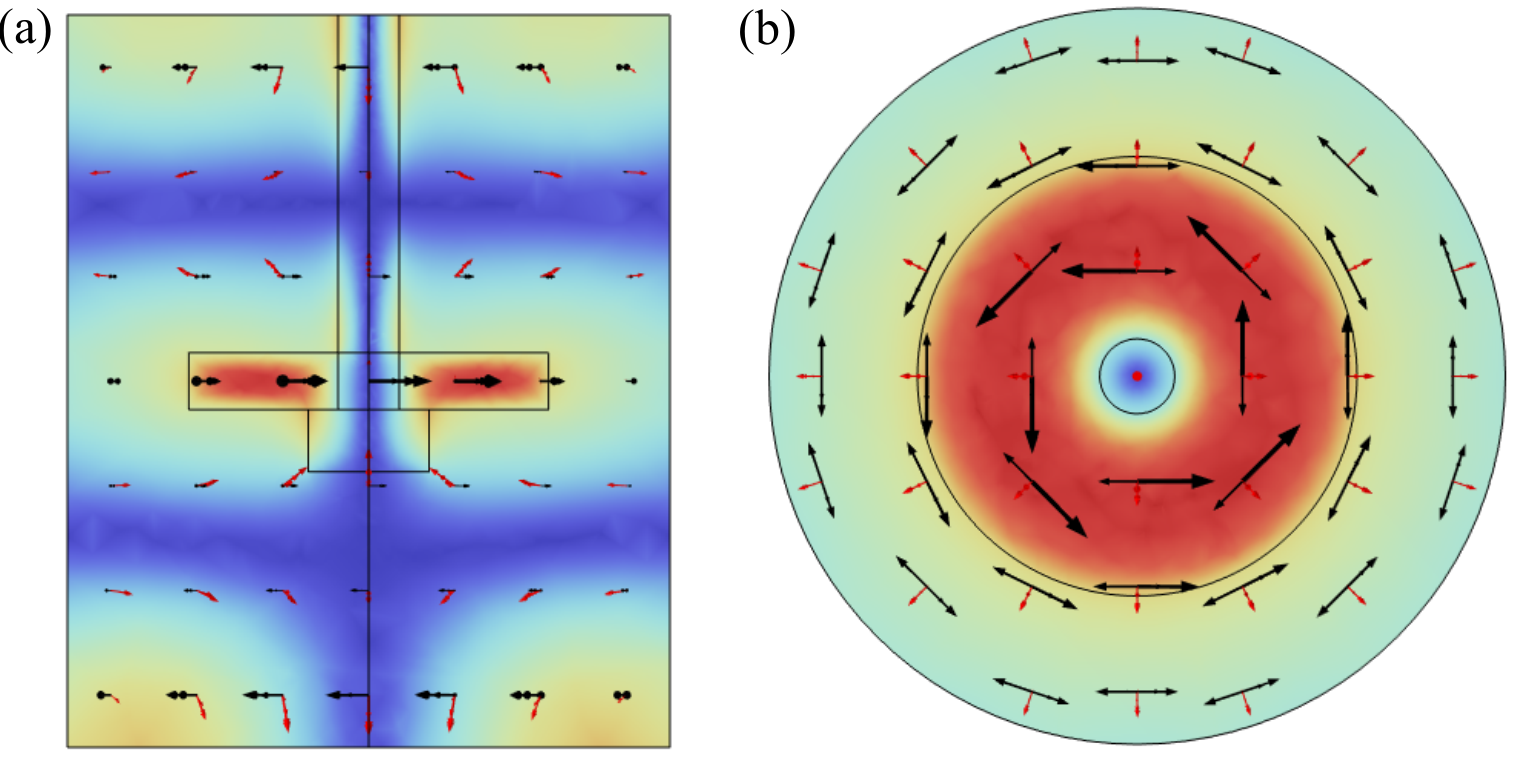}
\caption{Example of finite-element COMSOL simulation showing whispering-gallery quasi TM mode at 15.872~GHz with field distribution primarily inside 4H-SiC discs. Black arrows indicate the magnetic field; red arrows show the electric field at resonance. (a) XZ plane view. (b) XY plane view. (cavity geometry and dimensions used in the simulation are presented in the main text) }
\label{fig:microwave_modes}
\end{figure}

Several eigenmodes were identified in the frequency range of 10--22~GHz. WGMs are characterized by electromagnetic fields confined inside the 4H-SiC discs. Such modes exhibit high quality (Q) factors, enhancing spin-cavity coupling by reducing electromagnetic losses.
Figure \ref{fig:microwave_modes} presents the simulated field distribution of a WGM at 15.87~GHz SiC discs. Panels (a) and (b) show field confinement clearly in the XZ and XY planes, respectively.

Table \ref{tab:modes} summarizes selected simulated eigenmodes, indicating their frequencies and mode types (TM, TE, or WGM). Despite the simplified assumptions, these simulations confirmed the presence of WGMs and informed our experimental investigation.
While the exact experimental mode with the highest Q-factor did not directly correspond to any specific simulated mode, likely due to simplifying assumptions, the simulations nonetheless demonstrated the potential of 4H-SiC discs to support WGMs. 

\begin{table}
\centering
\caption{Eigenmodes calculated from FEM in the 4H-SiC Microwave Cavity (10--22~GHz).}
\label{tab:modes}
\begin{tabular}{|p{2cm}|p{2cm}|p{4cm}|}
\hline
\textbf{Mode} & \textbf{Frequency (GHz)} & \textbf{Type} \\
\hline
Mode 1 & 10.368 & TM \\
Mode 2 & 11.699 & TE \\ 
Mode 3 & 12.468 & TM \\
Mode 4 & 14.128 & WGM-TE (in 4H-SiC) \\
Mode 5 & 14.823 & TM \\ 
Mode 6 & 15.397 & WGM-TE (in 4H-SiC, $m=1$) \\
Mode 7 & 15.872 & WGM-TM (in 4H-SiC, $m=1$) \\
Mode 8 & 17.132 & TM \\
Mode 9 & 17.559 & TE \\ 
Mode 10 & 18.596 & WGM-TE (in 4H-SiC, $m=2$) \\
Mode 11 & 19.097 & TE \\ 
Mode 12 & 19.277 & WGM-TE (in sapphire) \\
Mode 13 & 19.281 & WGM-TE (in 4H-SiC, $m=3$) \\
Mode 14 & 20.293 & TM \\
Mode 15 & 20.628 & WGM-TM (in sapphire) \\
Mode 16 & 21.762 & TE \\ 
Mode 17 & 21.927 & WGM-TE (in 4H-SiC, higher order $m=2$) \\
Mode 18 & 22.000 & WGM-TE (in 4H-SiC, higher order $m=3$) \\
Mode 19 & 22.169 & WGM-TM (in 4H-SiC, $m=2$) \\
\hline
\end{tabular}
\end{table}

\section{Cavity mode and spin characterization at 4~K \label{AppendixB}}

At 4~K, we performed $S_{21}$ parameter microwave transmission measurements from 10~GHz to 22~GHz and identified five cavity mode resonances (Table~\ref{tab:measured_modes}); the mode with the highest quality (Q) factor appeared at 12.595~GHz. Q factors were extracted by fitting with Eq. (\ref{fitS21}) \cite{petersan_measurement_1998} and using $Q = f_0/\Delta f$. 

\begin{equation}
    |S_{21}(f)| = A_{1}+A_{2}f + \frac{S_{\max}+A_{3}f}{\sqrt{1+4[(f-f_{0})/\Delta f]^2}}
    \label{fitS21}
\end{equation}

\begin{figure}
   \centering
   \includegraphics[width=\linewidth]{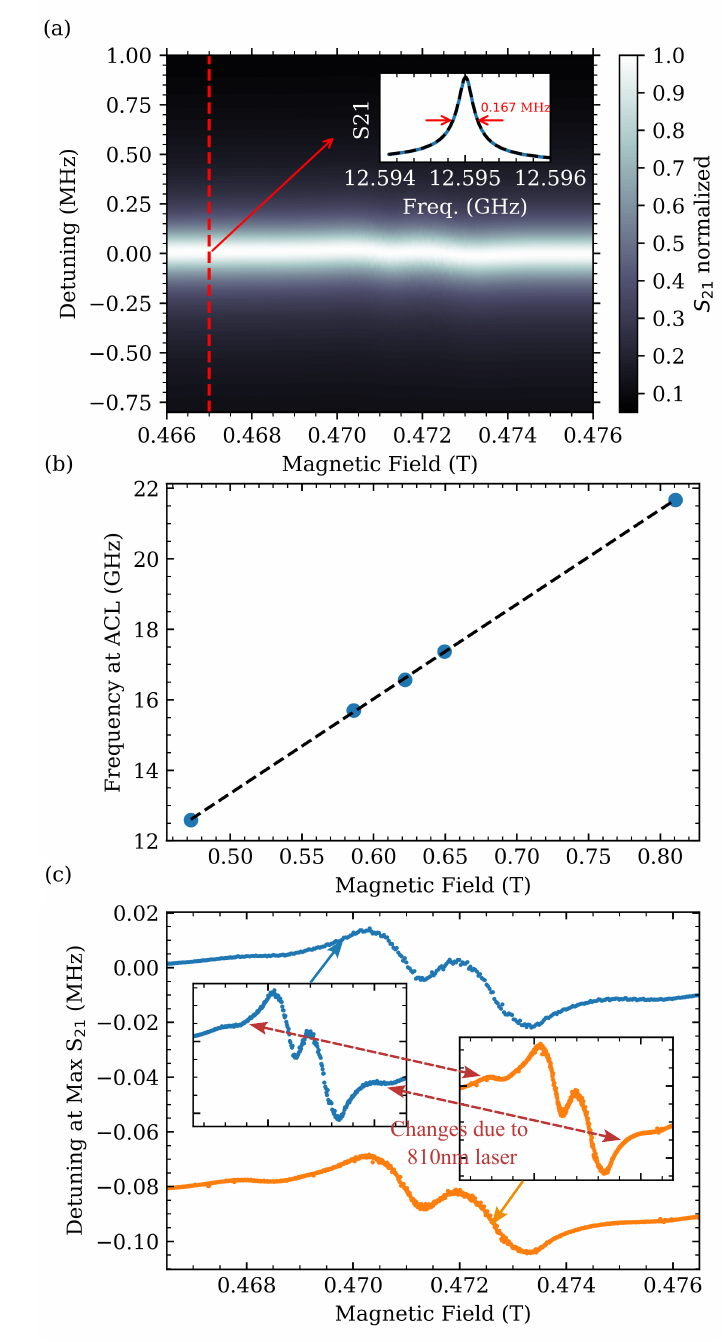}
   \caption{(a) Cavity transmission of the highest Q-factor mode found at 4~K temperature as a function of detuning from cavity resonance and axial magnetic field, revealing two avoided crossings corresponding to the two avoided crossings in the main text. Inset: S$_{21}$ cavity transmission far from spin resonances and fitted using Eq. (\ref{fitS21}) ($\omega_{0}/2\pi=12.595$~GHz, $\Delta \omega/2\pi =0.167$~MHz $Q\approx7.6\times10^{4}$). (b) Spin‐transition frequencies extracted from the five different avoided crossing lines (ACL) at the different cavity resonances listed in Table \ref{tab:measured_modes} and fitted with the spin-3/2 Hamiltonian Eq. (\ref{spinH}).The fitted values obtained were $\gamma_{e}/2\pi = 26.8\pm0.2$~GHz/T and $D/2\pi = 37\pm50$~MHz. (c) Detuning where S$_{21}$ is maximum as a function of magnetic field with and without optical excitation. Detuning is calculated with respect to cavity resonance at zero laser power. Blue curve measured with no laser excitation and orange curve measured using 810~nm laser excitation (for an arbitrary power, and we expect it to be of the order of a few mW). Curves are significantly offset due to the large optical excitation power since the cavity resonance changes with optical power. Similar to the results at 10~mK, we see changes in the avoided crossings of V$_2$ spin transitions as labelled by the red arrows. The spin-cavity interaction on the left becomes more prominent while the interaction on the right lessens.}
   \label{fig:4Kres}
\end{figure}

To examine spin-cavity interactions, we swept an axial magnetic field B$_{z}$ while measuring the transmission across each cavity mode. 
Figure~\ref{fig:4Kres}(a) shows two avoided crossing at the highest Q-mode at \(B\approx0.470\)~T and \(0.473\)~T, corresponding to the ground-state transitions of V$_1$ and V$_2$. 
Repeating the magnetic field sweep at the four lower-$Q$ modes revealed at most one avoided crossing per mode; the extracted measured spin transition frequencies are plotted in Fig.~\ref{fig:4Kres}(b).  
Fitting the data with the spin 3/2 Hamiltonian

\begin{equation}
    \hat{H}=D S_z^{2}+\gamma_e \mathbf{B}\cdot\mathbf{S}
    \label{spinH}
\end{equation}

gives an effective gyromagnetic ratio $\gamma_{e}/2\pi = 26.8\pm0.2$~GHz/T and $D/2\pi = 37\pm50$~MHz.
The uncertainty in $D$ is large because the two crossings are not resolved at every mode; however, the fitted $\gamma_{e}$ is within 5\% of the reported value for silicon vacancies $\gamma_{e}/2\pi = 28$~GHz/T (for $g_e=2.002$ \cite{son_ligand_2019}).  
The small deviation is attributed to either a misalignment of the SiC $c$-axis or a calibration offset in the SC magnet. A tilt between the discs and the magnetic field is sufficient to account for the observed shift. 

In Fig.~\ref{fig:4Kres}(c), the detuning from cavity resonance where S$_21$ transmission is maximum with and without optical excitation is plotted. The measured modes are significantly offset due to the large optical excitation power (on the order of mW) since the cavity resonance changes with optical power (see Appendix D). Similar to the measurements at 10~mK, there are small changes in the avoided crossings associated with V$_2$ spins. The overall avoided crossings remain similar in size, unlike the measurements at 10~mK, since laser heating at 4~K does not 
change the thermal Boltzmann distribution significantly.

We also note that we observe two avoided crossings in the middle instead of one feature as observed at 10~mK. The two middle features are also accompanied by two side features similar to the measurements at 10~mK. The two features on the side are consistent with the V$_2$ transitions observed at 10~mK. It is important to note that the SiC discs were repositioned before performing the measurements at 10~mK compared to the 4~K measurements, as both these measurements were performed in two separate sessions. The ZFS of spin-3/2 systems is known to vary with the relative angle between the applied magnetic field and the quantization axis \cite{gottscholl_superradiance_2022}. Therefore, the differences in the locations of the avoided crossings between 10~mK and 4~K measurements could be attributed to a change in the sample orientation, with the two middle features related to CAV$^+$ and V$_1$ spins and the side transitions related to $V_2$ spins.

\begin{table}
\centering
\caption{Cavity resonances measured at 4~K.}
\label{tab:measured_modes}
\begin{tabular}{|c|c|}
\hline
\textbf{Frequency (GHz)} & \textbf{$Q$ factor} \\
\hline
12.595 & $75\,526 \pm 4$ \\
15.697 & $779 \pm 2$ \\
16.574 & $5\,074 \pm 7$ \\
17.371 & $5\,790 \pm 12$ \\
21.674 & $31\,392 \pm 326$ \\
\hline
\end{tabular}
\end{table}

\section{Optical-spectroscopy of the SiC material \label{appendixC}}

\begin{figure}
  \centering
  \includegraphics[width=\linewidth]{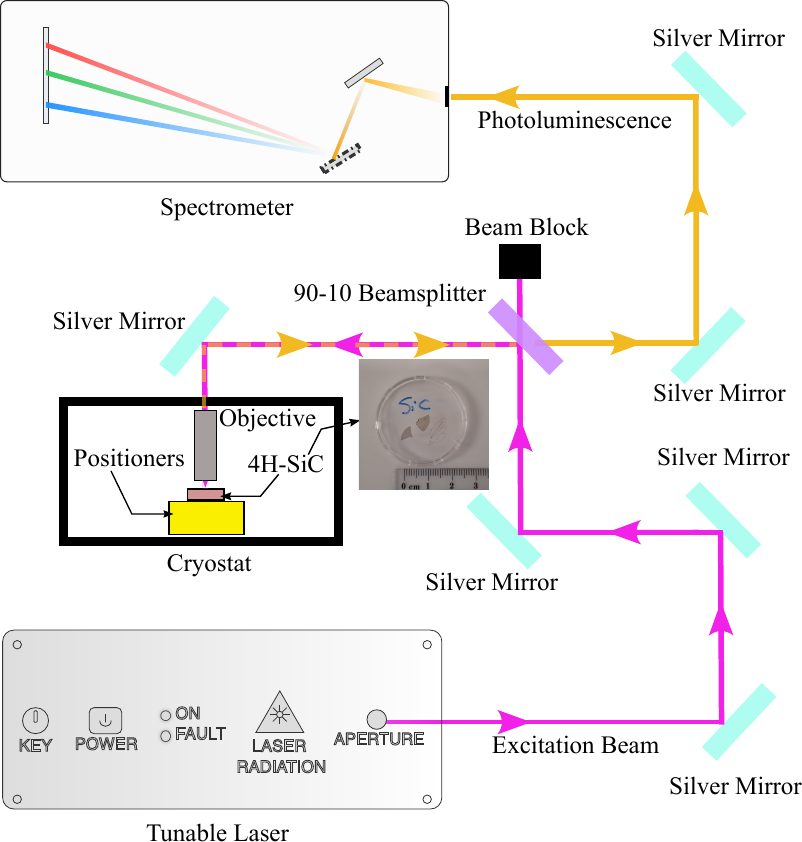}
  \caption{Schematic of the optical-spectroscopy setup. An image of the 4H-SiC pieces broken off from the full discs is shown and used separately for the optical spectroscopy.}
  \label{fig:setup_SI}
\end{figure}

The optical spectroscopy was performed in a confocal microcoscopy setup inside an attoDRY800 cryostat, and the schematic for the spectroscopy is shown in Figure \ref{fig:setup_SI}.
A tunable Ti:sapphire laser (M-Squared SolsTiS, 670–1050 nm) delivers 240 µW of 810 nm light to the sample (Fig.~S1).  Beam delivery employs protected-silver mirrors (Thorlabs PF10-03-P01) and a 90/10 beamsplitter (BS035).  The sample is mounted in an attoDRY800 cryostat equipped with a low-temperature apochromatic objective (NA 0.82) and XYZ nanopositioners (attocube ANPx311).  The emitted PL passes back through the beamsplitter and is dispersed by a Shamrock 750 spectrometer (Andor SOLIS).  Temperature is stabilized to ±0.1 K via the attoDRY800 system.

\begin{figure}
  \centering
  \includegraphics[width=\linewidth]{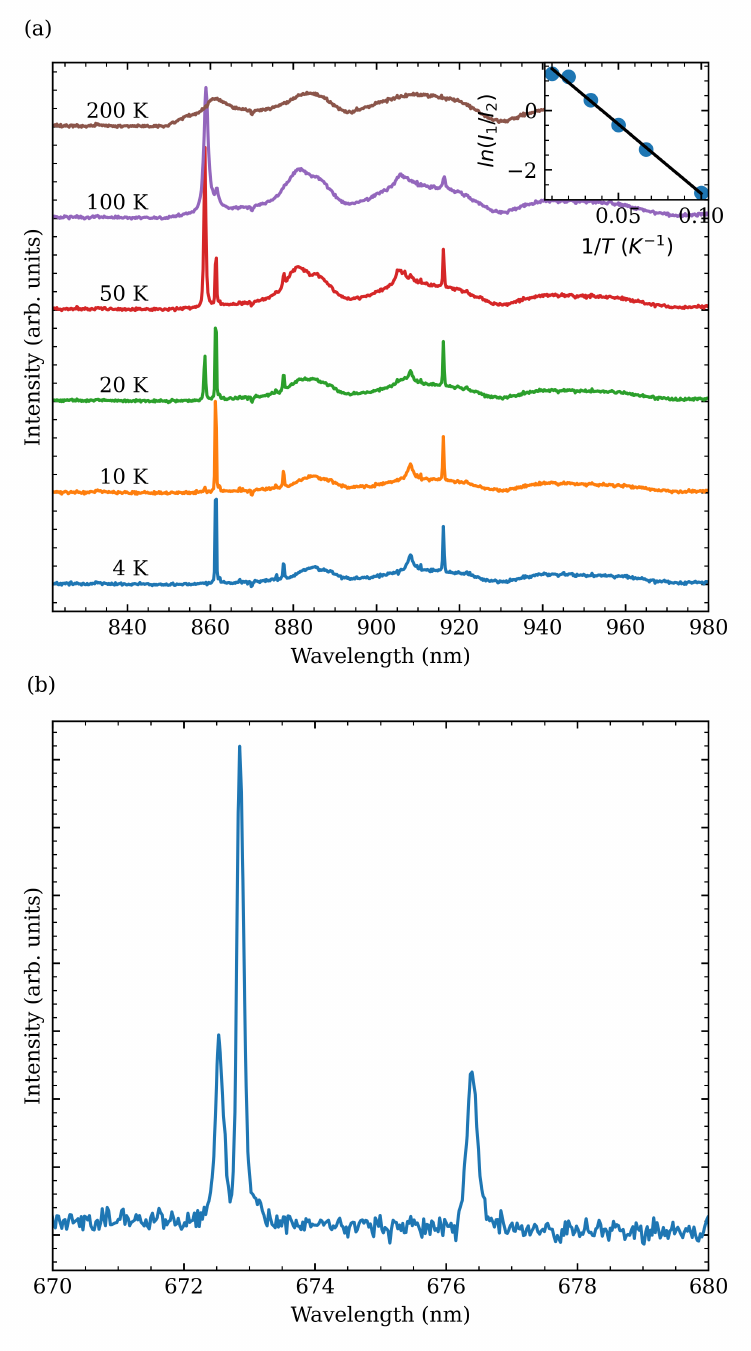}
  \caption{(a) Temperature-dependent PL spectra of 4H–SiC showing ZPL of V$_1$ (V$_1'$) at 861 (858)~nm and V$_2$ at 917~nm. Each temperature has been vertically offset for clarity. Inset: Plot of $I_{V1'}/I_{V1}$ vs $1/T$ fitted with linear line (fitted slope = $-46.77~ \pm~1.85~\text{K}=-4.03~\pm~0.10 ~\text{meV}$ consistent with the phonon mixing activation energy between V$_1$ and V$_1'$ excited states. \cite{nagy_quantum_2018}) (b) Emission spectrum of 4H-SiC under 532~nm optical excitation at 4~K showing ZPL of CAV$^+$ at 672.6~nm, 672.9~nm and 676.4~nm.}
  \label{fig:optical_spectroscopy}
\end{figure}

To verify the presence of silicon-vacancy spins in our 4H-SiC crystal, we performed photoluminescence (PL) spectroscopy between 4 and 200 K (see Fig.~\ref{fig:optical_spectroscopy}(a)).
At 4 K, three ZPLs are resolved at 858 nm (V${}_{1}^{'}$), 861 nm (V${}_{1}$), and 917 nm (V${}_{2}$).  
As temperature increases, the V${}_{1}$ ZPL quenches while V${}_{1}^{'}$ brightens, dominating the spectrum near 100 K; both vanish above 100 K. 
The V${}_{2}$ line is visible up to $\sim$50 K.
A linear fit to the plot of $\ln(I_{V_{1}'}/I_{V_{1}})$ versus $1/T$ yields an activation energy of $4.03\pm0.10$ meV, consistent with phonon-assisted excited state mixing of V${_1}$ and V${_1}'$ \cite{nagy_quantum_2018}, however, we do not examine these optical effects further.
Satellite peaks at 878 nm and 908 nm are assigned to modified V$_{\text{Si}}$ (V${}_{\text{Si}}^{-}$ defects perturbed by carbon antisites) \cite{davidsson_exhaustive_2022}. These spectra confirm that both V${}_{1}$ and V${}_{2}$ species are optically addressable at cryogenic temperatures. 

In addition, optical spectroscopy was also performed with a 532~nm laser (OBIS LS) with a power of 100~$\mu$W and a 650 long pass filter (Thorlabs FELH-650) was used to filter the laser light before detection of sample fluorescence.
Using 532nm incident light on the sample, we collect emission from 670~nm to 680~nm (see Fig. \ref{fig:optical_spectroscopy}(b)) as this is the range we expect spin-1/2 CAV$^+$ defects that could influence our microwave measurements (i.e. will contain spins with frequencies proportional to $\gamma_e B$). We detect ZPL lines B$_1 =672.6~$nm, B$_2 =672.9~$nm and B$_3 =676.4~$nm. There is usually an additional peak near B$_3$ reported in the literature \cite{castelletto_silicon_2014}, which we cannot clearly distinguish in our measurements.

\section{Effect of optical pump power on cavity modes}
\begin{figure*}
  \centering
  \includegraphics[width=\linewidth]{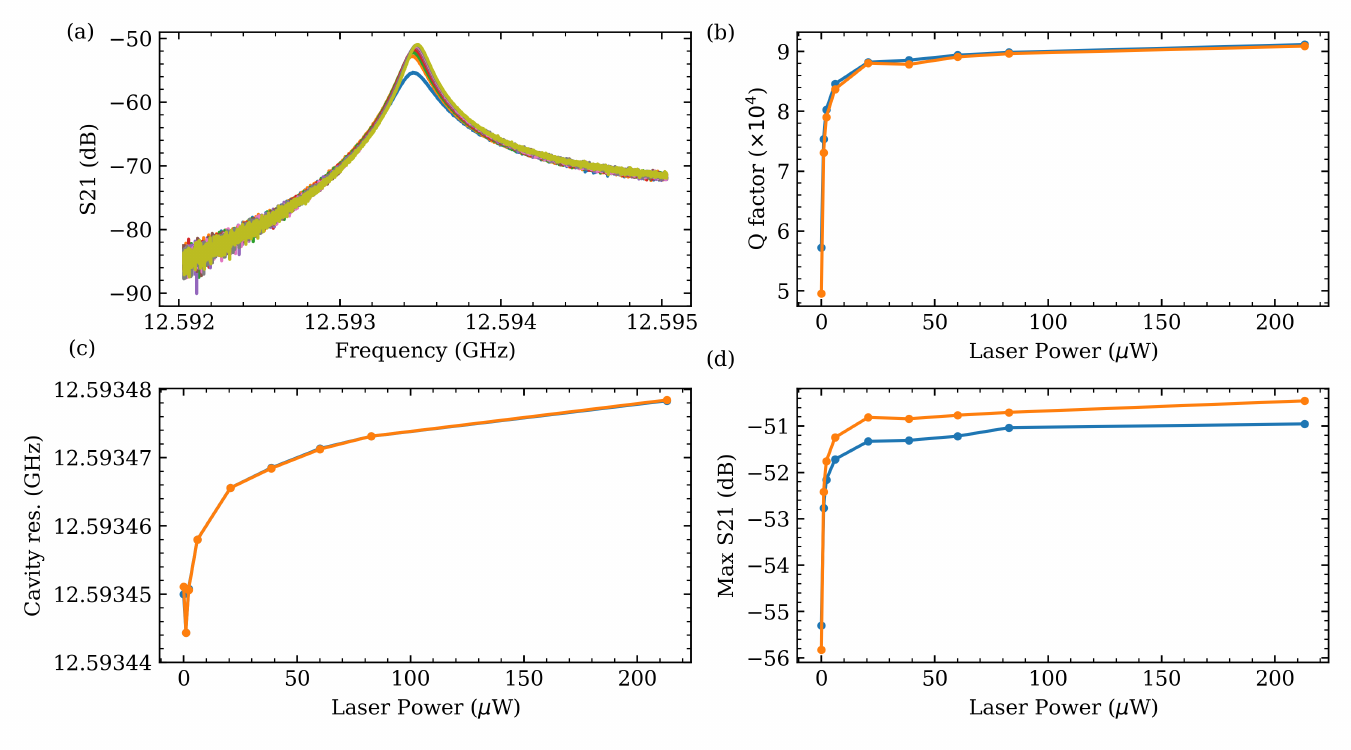}
  \caption{ Effect of optical pump power at 810~nm on the microwave cavity mode at zero magnetic field. (a) $S_{21}$ cavity transmission spectra measured at different optical pump powers  (blue:~0~$\mu$W, orange:~0.9~$\mu$W, green:~2.1~$\mu$W, red:~6.0~$\mu$W, purple:~20.6~$\mu$W, brown:~38.6~$\mu$W, pink:~60.1~$\mu$W, grey:~82.7~$\mu$W and olive:~213.0~$\mu$W). (b) Cavity $Q$ factor as a function of optical pump power. Blue curve measured using -20~dBm and orange curve using -30~dBm microwave power. (c) Cavity resonance frequency as a function of optical pump power. (d) $S_{21}$ maximum amplitude as a function of optical pump power}
  \label{fig:optical_power_effect}
\end{figure*}

We characterized the impact of optical pumping at 810~nm on the cavity mode properties by measuring the $S_{21}$ transmission spectra at different pump powers, with no magnetic field applied. 

Figure~\ref{fig:optical_power_effect}(a) displays representative cavity transmission spectra measured at selected optical powers, clearly showing that the resonance frequency slightly shifts with increased pump power.

The cavity quality factor $Q$ was extracted from fits to the transmission spectra as detailed earlier in Eq. (\ref{fitS21}), and is plotted against optical power in Fig.~\ref{fig:optical_power_effect}(b). We observe a clear increase in the cavity $Q$ factor as the optical power increases, suggesting that optical pumping reduces internal microwave losses.

Figure~\ref{fig:optical_power_effect}(c) summarizes the optical power-dependent shift in cavity resonance frequency. Far from spin resonance, the cavity resonance frequency increases with increasing laser power. This trend suggests that optical absorption at low temperature leads primarily to a change either in the dielectric environment \cite{hartnett_microwave_2011} or cavity geometry that shifts the cavity resonance upward. However, this behavior contrasts with Fig. \ref{fig:discussion}(a) and Fig. \ref{fig:4Kres}(c), where near spin resonance the cavity resonance decreases with increasing optical power. A similar decrease is also observed in Fig. \ref{fig:mwpower}(c) when the microwave power is increased near resonance. These results indicate that, in the vicinity of spin resonance, additional mechanisms, likely related to microwave driving of the spin transitions, lead to a net downward shift of the cavity resonance frequency. 

Importantly, while these static cavity properties show clear optical power dependence, the fitting of the measured data obtained with optical pumping was performed using the optically modified cavity resonance frequencies measured in this section.

\section{Thermal effects on spin-photon coupling strength}

From the measured coupling reduction at the avoided crossing at B$_2$ in the main text ($g_2^{(\text{opt})}/g_2^{(\text{dark})} = 0.30$), we extract the effective spin temperature for a spin-$3/2$ system since this transition corresponds to V$_2$ spins. We assume that this reduction is purely due to the thermal redistribution of spin populations, consistent with Boltzmann statistics. For spin-3/2 systems, the temperature-dependent coupling ratio can be estimated with the following expression using $g\propto \sqrt{N}$ \cite{amsuss_cavity_2011} for N spins and at higher temperatures $g\propto\sqrt{\Delta N}$ where $\Delta N$ is the population difference between the cavity-coupled spin states since the effective number of spins is reduced due to the Boltzmann population \cite{Miyashita_2012}.

Solving Eq. (\ref{eq:ratiog}) for the temperature T gives 1.7625~K. At $T = 1.7625$~K, the distributed populations of the four spin levels m$_s=\{-3/2,-1/2,1/2,3/2\}$ are $\{0.32044,\ 0.22738,\ 0.22652,\ 0.22566\}$. 

\begin{multline}
    \left(\frac{g_{S=3/2}(T)}{g_{S=3/2}(10\text{mK})}\right)^2 \approx \\
    \frac{1-e^{-\hbar \omega_c/k_BT}}{1+e^{-\hbar \omega_c/k_BT}+e^{-(\hbar \omega_c+2D_i)/k_BT}+e^{-(\hbar \omega_c+4D_i)/k_BT}} 
    \label{eq:ratiog}
\end{multline}

The population difference between the first two levels ($\Delta N_{1-2} \approx 0.09306$) is much larger than between the upper two levels ($\Delta N_{3-4} \approx 0.00086$). Under a purely thermal distribution, the avoided crossing at B$_3$ (which involves the upper levels) should therefore have a coupling much smaller than the first avoided crossing. 

Experimentally, however, the fitted coupling strengths $g_2 = 0.41$~MHz and $g_3=0.37$~MHz for the avoided crossings at B$_2$ and B$_3$, respectively, are similar. This indicates that the upper spin states exhibit a greater population difference than expected by Boltzmann statistics alone, implying an additional mechanism, which we discuss as spin-selective optical pumping in the main text. 

\section{Effect of microwave power on spin-cavity modes}

\begin{figure*}
    \centering
    \includegraphics[width=\linewidth]{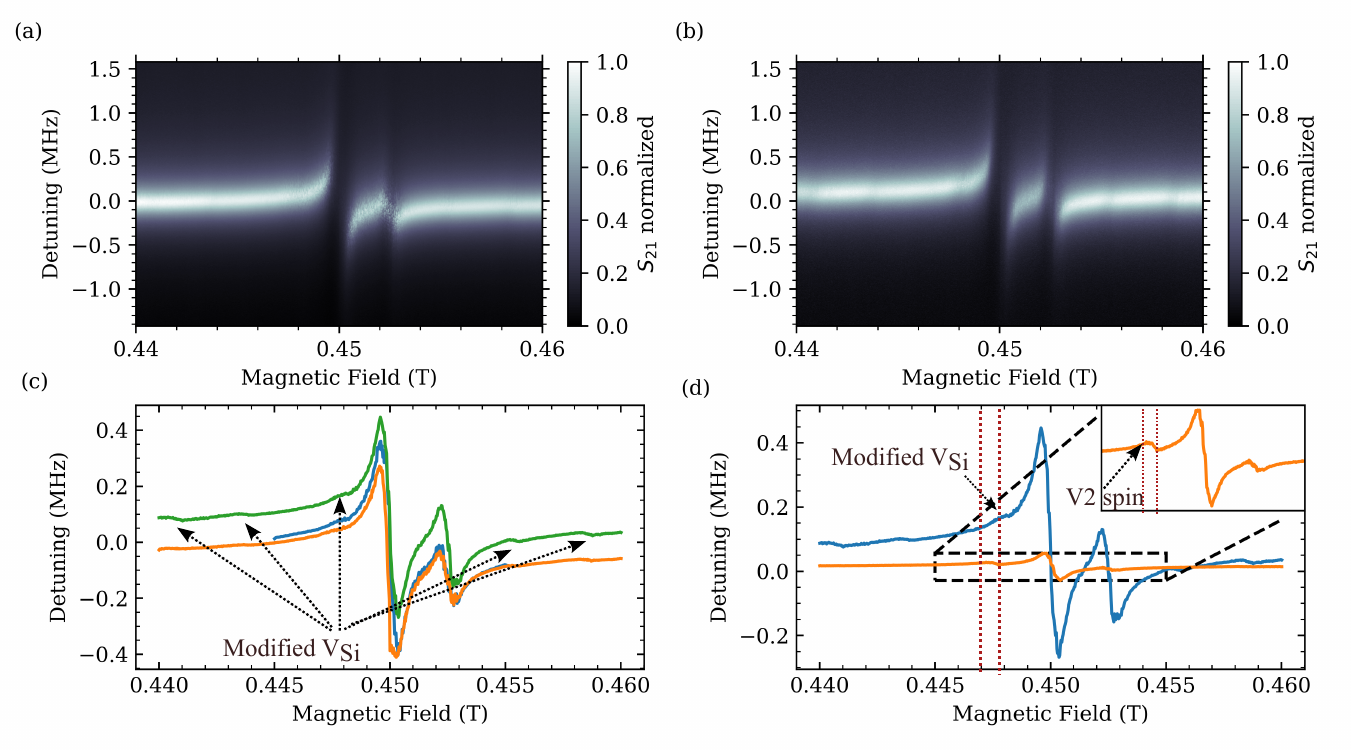}
\caption{Effect of microwave power at 10~mK: (a) S$_{21}$ cavity transmission plot measured using -20~dBm microwave input power (identical to the power used for the results obtained in the main text), however, here the data was obtained using an IFBW=50~Hz on the network analyser instead of IFBW=10~Hz, which was used for the results in the main text and a larger magnetic field range is covered. (b) S$_{21}$ cavity transmission using -30~dBm microwave input power. At lower power, coupling to modified V$_{Si}$ defects is more visible. (c) The frequency of the hybrid modes obtained by fitting the $S_{21}$ with Eq. (\ref{fitS21}). The blue curve is the data presented in the main text, which was obtained using -20~dBm microwave power and IFBW=10~Hz; the orange curve is the data obtained over a larger magnetic field range using -20~dBm and IFBW=50~Hz; and the green curve is obtained at -30~dBm using IFBW=50~Hz. A vertical shift is observed between the curves of different microwave powers. Additional features which we attribute to modified V$_{Si}$ defects are labelled with arrows. (d) The modified V$_{Si}$ seen in the 10~mK measurement with no optical pumping (blue) is close to the third feature observed with optical pumping (orange) (see main text for more details on this additional feature that arises when optical pumping is present); however, they occur at different magnetic fields (as shown by the vertical red-dashed lines) and therefore they are not the same feature. Inset: shows a zoomed-in plot of data obtained under optical pumping.}
    \label{fig:mwpower}
\end{figure*}

We use two different microwave powers as well as different intermediate frequency bandwidth (IFBW) settings on the network analyser that was used to take the $S_{21}$ measurements. 
In Fig. \ref{fig:mwpower}(a), we show the S$_{21}$ cavity transmission measured at 10~mK using a microwave input power of -20~dBm and an IFBW of 50~Hz, in contrast to the narrower IFBW of 10~Hz used for the data presented in the main text. The wider IFBW significantly reduces the measurement time at each magnetic field point, enabling more efficient scanning across a broader magnetic field range. In Fig. \ref{fig:mwpower}(a), the S$_{21}$ transmission is shown for a lower input power of -30~dBm. At this reduced power, coupling to additional spins becomes more apparent.

Figure \ref{fig:mwpower}(c) shows the central frequencies of the hybrid modes obtained by fitting the S$_{21}$ transmission spectra with Eq. (\ref{fitS21}). The blue curve corresponds to the data from the main text (-20~dBm, IFBW=10~Hz), the orange curve shows data acquired over the extended magnetic field range with the same power but a wider IFBW of 50~Hz, and the green curve represents data collected at -30~dBm with IFBW=50~Hz. A vertical shift is observed between the curves taken at different microwave powers. Interestingly, as the microwave power is increased from -30~dBm to -20~dBm, the cavity resonance frequency reduces. This is the opposite effect to what was observed with increasing optical power, where the resonance frequency increases with optical power (see Fig. \ref{fig:optical_power_effect}). Notably, several additional avoided crossing features are visible and are highlighted with arrows. These features do not correspond to known hyperfine transitions associated with $^{13}$C or $^{29}$Si \cite{son_ligand_2019}, and we attribute them to modified negatively charged V$_{\text{Si}}$ defects. Their appearance is consistent with theoretical simulations of the zero-field splitting parameter $D$ for these modified defects, which predict a range including but not limited to $D\approx-400 \text{ to } 80$~MHz \cite{davidsson_exhaustive_2022}.

In Fig. \ref{fig:mwpower}(d), we compare the transmission spectra obtained at 10~mK in the absence of optical pumping (blue) with those obtained under optical pumping (810nm, 20~$\mu$W) conditions (orange). One of the modified V$_{\text{Si}}$ sites observed without optical pumping appears near the third feature reported in the main text; however, the two occur at different magnetic field values and therefore arise from distinct spin transitions. In addition, under optical pumping, the coupling of modified V$_\text{Si}$ sites is no longer observed due to thermal effects. The inset shows a zoomed-in view of the data acquired under optical pumping, highlighting this distinction more clearly.

\clearpage
\bibliography{references.bib}

\end{document}